\documentclass{PoS}
\usepackage{amsmath,epsfig,cite,graphics}
\usepackage{macros_static}
\usepackage{wrapfig}

\title{Preparing for $N_f=2$ simulations at small lattice spacings}

\ShortTitle{$N_f=2$ simulations at small lattice spacings}

\author{\LALPHA \hfill
        \onecol{4.0cm}{\vspace{-1.5cm}\it DESY 07-159 \\ SFB/CPP-07-55 
          \\MS-TP-07-35\\CERN-PH-TH/2007-170 \\ HU-EP-07/46 \\ MIT-CTP 3873
          \vspace{-2.1cm}
        }}
\author{M. Della Morte,\\
        CERN, Physics Department, TH Unit,
        CH-1211 Geneva~23, Switzerland
       }
\author{P. Fritzsch,\\
        University of M\"unster,
        Wilhelm-Klemm-Strasse~9, D-48149 M\"unster, Germany
       }
\author{B. Leder, S. Takeda, O. Witzel, U. Wolff\\
        Humboldt University, Newtonstr.~15, 12489 Berlin, Germany\\
       }
\author{H. Meyer,\\
        MIT, Cambridge, MA 02139, U.S.A.
        \\
        }
\author{H. Simma\thanks{
        present address: University Milano Bicocca, Pz. della Scienza 3,
        20126 Milano, Italy.
        }, \speaker{R. Sommer}%\thanks{A footnote may follow.}
        \\
        DESY, Platanenalle 6, 15738 Zeuthen, Germany\\
        E-mail: \email{rainer.sommer@desy.de}
        }

\abstract{We discuss some large effects of dynamical fermions. One is a
cutoff effect, others concern the contribution of
multi-pion states to correlation functions and are
expected to survive the continuum limit.
We then turn to the  preparation for simulations at small
lattice spacings which we are planning down to around
$a=0.04~\fm$ in order to understand the size
of O($a^2$)-effects of the standard O($a$)-improved theory.
The dependence of the lattice spacing
on the bare coupling is determined through the Schr\"odinger
functional renormalized coupling.
}

\FullConference{The XXV International Symposium on Lattice Field Theory\\
		 July 30-4 August 2007\\
		 Regensburg, Germany}

\begin{document}

\section{Introduction}
The ALPHA collaboration has worked over the years on a determination
of the QCD $\Lambda$-parameter starting from experimental low energy
hadronic input and using perturbation theory in a renormalized 
coupling at sufficiently
high energy scales. At these scales it was demonstrated that
perturbation theory is very accurate. The quoted results for the $\msbar$ 
$\Lambda$-parameter are $\Lambda_\msbar^{(0)}r_0 = 0.60(5)$ \cite{mbar:pap1}
in the quenched approximation
and $\Lambda_\msbar^{(2)}r_0 = 0.62(4)(4)$ \cite{alpha:nf2}
with $\nf=2$ dynamical quarks. 
On the other hand the $\nf=5$ value extracted by matching various 
experimental data to perturbation theory in the (not always very) high
energy region translates into $\Lambda_\msbar^{(5)}r_0 \approx 0.55$ 
\cite{alpha:bethke06},
when $r_0=0.5\,\fm$ \cite{pot:r0} is used. Superficially this suggests a nice
agreement, but the perturbative matching across the quark thresholds
\cite{alpha:bernwetz} 
yields
$\Lambda_\msbar^{(4)} / \Lambda_\msbar^{(5)} \approx 1.4$ which does not
connect smoothly to the $\nf=0,2$ numbers. In order to say more about 
this comparison, the
low energy scale $r_0$ should be replaced by an experimental observable
and the continuum limit should be evaluated with a better confidence
than it was possible in \cite{alpha:nf2}. 
%% Here we discuss the reasons 
%% for the latter statement as well as a first step, namely the determination
%% of the simulation parameters. 
Significant progress in the understanding of the 
continuum limit requires to simulate smaller lattice spacings
with good accuracy.
We will motivate this further in \sect{s:aeffects}. The difficult
simulations are the ones in large volume where for example the Kaon decay
constant is to be determined to set the energy scale in $\GeV$. 
We will briefly explain in
\sect{s:SF} that our previous approach of using \SF boundary conditions
in that part of the calculation meets somewhat unexpected
(practical) difficulties. Since these are related to true dynamical 
fermion effects,
they are theoretically interesting, but it appears to be better to switch to
(anti)-periodic boundary conditions in this part of the calculation.
In \sect{s:lstar} we will finally discuss a determination of the 
dependence of the lattice spacing on the bare coupling. This represents
a useful piece of information for fixing the parameters of the large volume
simulations.

Before entering our discussion we add a comment on the motivation. One
might object to the whole project of a determination of the $\Lambda$-prameter 
that 
very precise lattice determinations for 
$\alpha_\msbar(M_\mrm{Z})$ have already been published
\cite{alpha:lepagemilk}. However, apart from the use of rooted
staggered fermions, in these determinations perturbation theory 
has been used at rather low renormalization scales and for non-universal
quantities (small Wilson loops). These are defined at the scale 
of the (lattice)-cutoff. It is
apparent from the discussion in \cite{alpha:lepagemilk} that the use of 
perturbation theory is problematic. The known terms in the expansion
either have large coefficients or, if one resums by choosing a different
scheme, the renormalization scale becomes even smaller and the expansion
parameter larger. In order to describe
the data several higher order terms in the expansion are fitted. 
Thus it appears that a computation following the ALPHA-strategy, where
the continuum limit is taken and 
perturbation theory is verified to apply for the considered 
{\em renormalized} coupling, 
remains very well motivated. We do not see any alternative to this strategy 
if a full control of all systematics is desired.

In the following considerations we use the standard O($a$)-improved theory
with Wilson's gauge action and the non-perturbatively determined~\cite{impr:csw_nf2} 
coefficient $\csw$ of the Sheikholeslami-Wohlert term \cite{impr:SW}.

%%%%%%%%%%%%%%%%%%%%%%%%%%%%%%%%%%%%%%%%%%%%%%%%%%%%%%%%%%%%%%%%%%
\section{Cutoff effects in $\za$ \label{s:aeffects}}
In \cite{lat03:rainer} we have presented evidence that cutoff effects
tend to be larger in full QCD than they are   
in the quenched approximation. Here we would like to draw the reader's 
attention to the non-perturbative 
determination of $\za$ presented in \cite{impr:za_nf2}.
It uses a Ward identity in the \SF in an $L^3\times9/4L$ geometry with
$L\approx 0.8\,\fm$ as in the quenched approximation \cite{impr:pap4}. 
It can be shown that the quark-propagator disconnected diagrams which
enter the Ward idenity vanish in the continuum limit. They  
are of $\rmO(a^2)$ at a finite lattice spacing. In contrast to the 
quenched approximation where already at $a=0.1\,\fm$ they were insignificant 
(in 
comparison to the numerical precision), for 
$\nf=2$ they 
contribute an about 15\% effect in $\za$ at such a lattice spacing. 
Even if this effect disappears very quickly at smaller $a$, it is
unpleasantly large at the lattice spacings one typically would like 
to include in a continuum extrapolation. 

In the mean time we have investigated the problem further, finding
that qualitatively this effect persists if one changes the 
angle $\theta$ in the spatial fermion boundary condition. 
Alternatively we considered the Ward identity between static-light states 
in such a way that disconnected diagrams are absent. Unfortunately, even when using
HYP discretizations for static quarks \cite{stat:actpaper} 
the statistical errors in $\za$ become relatively 
large at the smaller lattice spacings. Still, we confirmed that 
$\za$ defined in this way is rather close to the definition with
light-light states but disconnected diagrams dropped.

In general, cutoff effects are expected to be more prominent
in correlation functions (and for time separations) where excited 
state contributions are very important. We therefore investigate
at present whether the approximate ground state projection 
of \cite{impr:ca_nf2} suppresses the disconnected contribution
to $\za$ thus accelerating the continuum limit. Whether this attempt 
is successfull or not, these
difficulties suggest that one most likely needs smaller $a$ with 
dynamical fermions than in the quenched approximation. 
We now turn to another strong effect of dynamical fermions -- one that
is expected to persist in the continuum limit.

%%%%%%%%%%%%%%%%%%%%%%%%%%%%%%%%%%%%%%%%%%%%%%%%%%%%%%%%%%%%%%%%%%
\section{The large-volume \SF  \label{s:SF}}

\begin{figure}[t]
\vspace{-0.0cm}
\begin{center}
\includegraphics[scale=0.55]{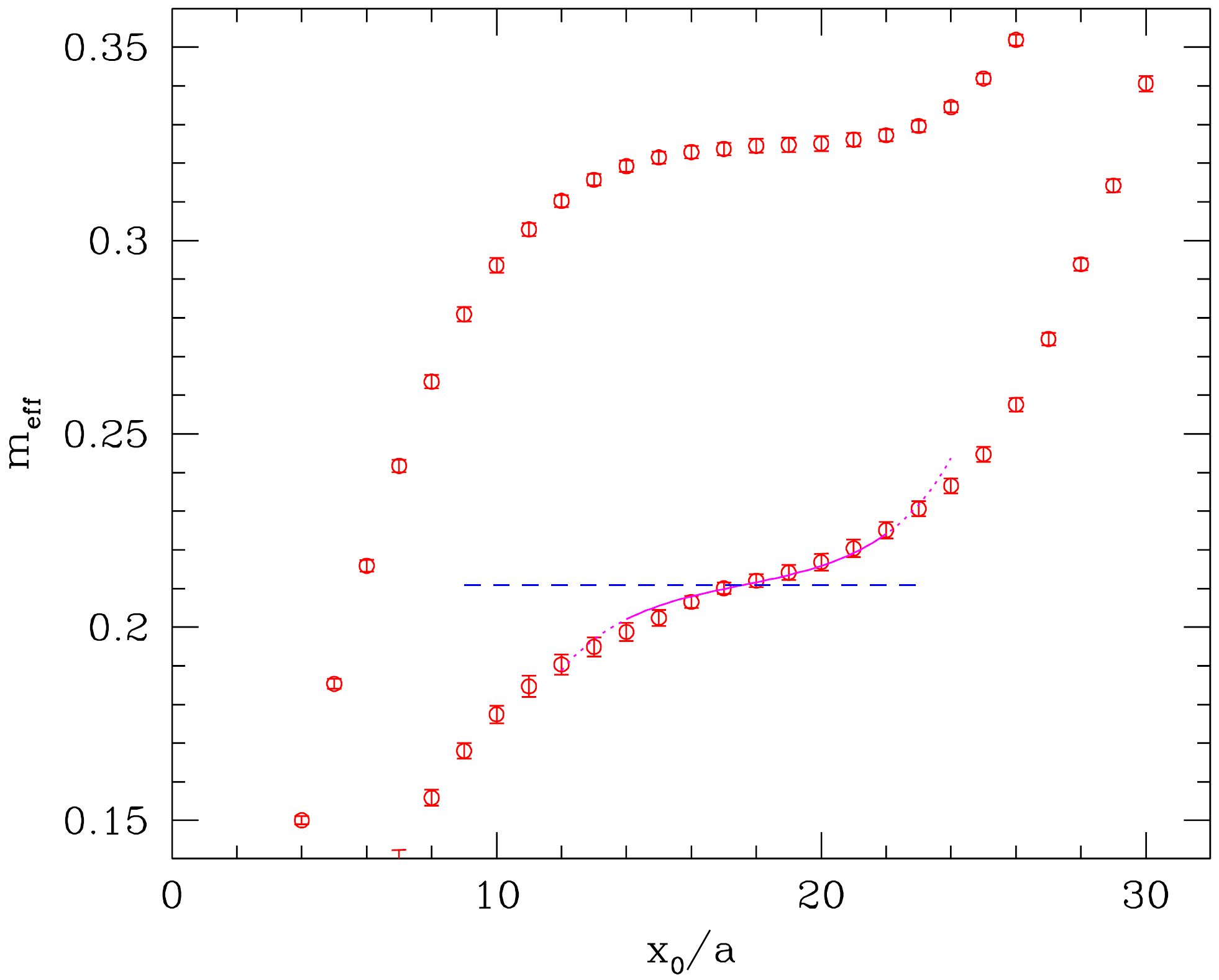}
\end{center}
\vspace{-0.5cm}
\caption[]{\label{f:plateaux}
The effective mass for the \SF correlation function 
$\fp$ at $\beta=5.3$ on a $24^3\times 32$ lattice
for $\kappa=0.1355$ and $\kappa=0.13605$. The fit described in
the text is extended outside the fit-range as a dotted curve.
The dashed line indicates the fitted pion mass. 
}
\end{figure}

Apart from the non-perturbative evaluation of renormalization
constants, the \SF also proved to be advantageous  for the computation 
of hadron masses and matrix elements such as $\Fk$ in the quenched 
approximation\cite{mbar:pap2}. A time extent of $T=3\,\fm$ allowed to clearly 
isolate ground state contributions. 
We have then
attempted to compute the pseudoscalar masses and decay constants
for $\nf=2$ with an $L^3 \times T$ Schr\"odinger functional, 
keeping $L\geq2 \,\fm$ 
and $T\approx 2.5\,\fm$. Indeed, at a quark mass around the
physical strange quark mass ($\kappa=0.1355$), the effective mass of
the pseudoscalar correlation
functions $\fa,\fp$ (see e.g. \cite{mbar:pap4} for their definition) 
exhibit short plateaux. An example is shown in the upper part of
\fig{f:plateaux}.  
%%%%%%%%%%%%%%%%%%%%%%%%%%%%%%%%%%%%%%%%%%%%%%%%%%%%%%%%%%%%
%%%%%%%%%%%%%%%%%%%%%%%%%%%%%%%%%%%%%%%%%%%%%%%%%%%%%%%%%%%%

However, the plateaux disappear quickly when the quark mass 
is lowered. For a quark mass of about half the strange quark mass
($\kappa=0.13605$),
excited state contaminations are strongly present in both 
the vacuum channel and in the pion channel. The former yield
contributions $\propto \exp(-(T-x_0)\,E^\mrm{vac}_{1})$ and
the latter $\propto \exp(-x_0\,(E^\mrm{\pi}_{1}-\mpi))$. Once 
these two leading contaminations are included, fits to
the correlation functions are still reasonable. We show a fit where
we have fixed $E^\mrm{vac}_{1} = 2\mpi$, $ E^\mrm{\pi}_{1} = 3\mpi$. These
are the energies of the multi pion states with the correct quantum 
numbers, when the interaction of the pions is neglected. 
At sufficiently large $L$ this is a good approximation.

We may conclude that {\em multi-pion states are observed}, as expected 
in the full theory. Their amplitude appears to be significantly 
stronger than with (point-to-point correlators and) 
periodic boundary conditions \cite{cern:I}. The standard \SF 
boundary operators have a strong overlap with these states. 
Even though it is interesting to observe these {\em strong effects 
of dynamical fermions} and a consistent description over a significant
range of $x_0$ can be achieved in the form of a fit, their presence 
hampers a reliable estimation of the systematic errors. We have hence 
decided to switch to
periodic boundary conditions for the purpose of computing large
volume matrix elements.

%%%%%%%%%%%%%%%%%%%%%%%%%%%%%%%%%%%%%%%%%%%%%%%%%%%%%%%%%%%%%%%%%%
\section{The lattice spacing as a function of the bare 
  coupling  \label{s:lstar}}

As a first step towards such computations we now compute,
in a massless renormalization scheme, the dependence
$a(g_0)$
of the lattice spacing on the bare coupling $g_0$ for 
$0.04\,\fm\; \lessim\; a \; \lessim \;0.1\,\fm$. 
Of course the function 
$a(g_0)$ is not unique, but only defined up to cutoff effects,
which depend on the renormalized quantity that is held fixed.
We employ a renormalization condition
which is relatively easily evaluated and which does not
introduce artificially large $a$-effects. This has proven to be 
the case for the standard
\SF coupling $\gbar^2(L)$, defined in \cite{alpha:su3,pert:1loop},
at vanishing quark mass. 

We further specify a scale $L^*$ by
\bes \label{e:lstar}
    \gbar^2(L^*) = 5.5 \,,
\ees
which is known to lead to $L^*/a\, \grtsim\, 8$ for the planned range of $a$. 
For such a choice, table~7 of \cite{alpha:nf2} shows a
change of $\gbar^2$ by about $\Delta \gbar^2 = 0.3$ when the 
boundary $\Oa$ improvement  coefficient $\ct$ is changed from
its 1-loop to its 2-loop approximation. Using the non-perturbative
beta-function of \cite{alpha:nf2}, 
\bes
  \label{e:ldep}
  L {{\rm d} \over {\rm d} L} \gbar^2 = -2\,\gbar\,
  \beta(\gbar) = 0.21(1)\,\gbar^4 \; \;\text{at }
  \gbar^2\approx 5.5\,,
\ees
a value $\Delta \gbar^2 = 0.3$ translates into a 5\% change in
 $L^*$ and thus $a$. 
The definition \eq{e:lstar} is completed by an exact definition
of the massless point. We choose the
PCAC mass $m$ (with non-perturbative $\ca$ \cite{impr:ca_nf2})
with \SF boundary conditions, with
$
    T=L=L^*\,,\; \theta=0.5
$
and a vanishing background field.
\begin{table}[t]
 \centering
  \begin{tabular}{llllll}
   \hline\\[-1.0ex]
   $L/a$ & $\beta$ & $\kappa$ & $\bar{g}^2(L)$ & $am$
\\[1.0ex]
   \hline\\[-1.0ex]
8   & 5.3    &0.136197  &5.65(5) & $\phantom{-}0$ \\
8   & 5.3574 &0.13564  &5.59(5) &  $\phantom{-}0.024(1)$ & \\
8   & 5.3574 &0.1367  &4.98(13) &  $-0.011(1)$ & \cite{alpha:nf2}\\
8   & 5.3574 &0.136365  &5.26(6) & $\phantom{-}0$ &interpolated \\
10  & 5.5    &0.136712  &5.11(8) & $-0.0008(2)$ \\
12  & 5.6215 &0.136665  &5.62(9) & $\phantom{-}0.0019(2)$ \\
16  & 5.8097 &0.1366077 &5.48(12)& $\phantom{-}0$ &\cite{alpha:nf2} \\[1.0ex]
   \hline
  \end{tabular}
 \caption{Raw simulation results and interpolated values. Values
 of $am=0$ indicate that $|z|=|Lm|$ is estimated to be at most $5\times10^{-3}$. }
 \label{tab:raw}
\end{table}

\begin{figure}[tb]
\begin{center}
\includegraphics*[width=10.5cm]{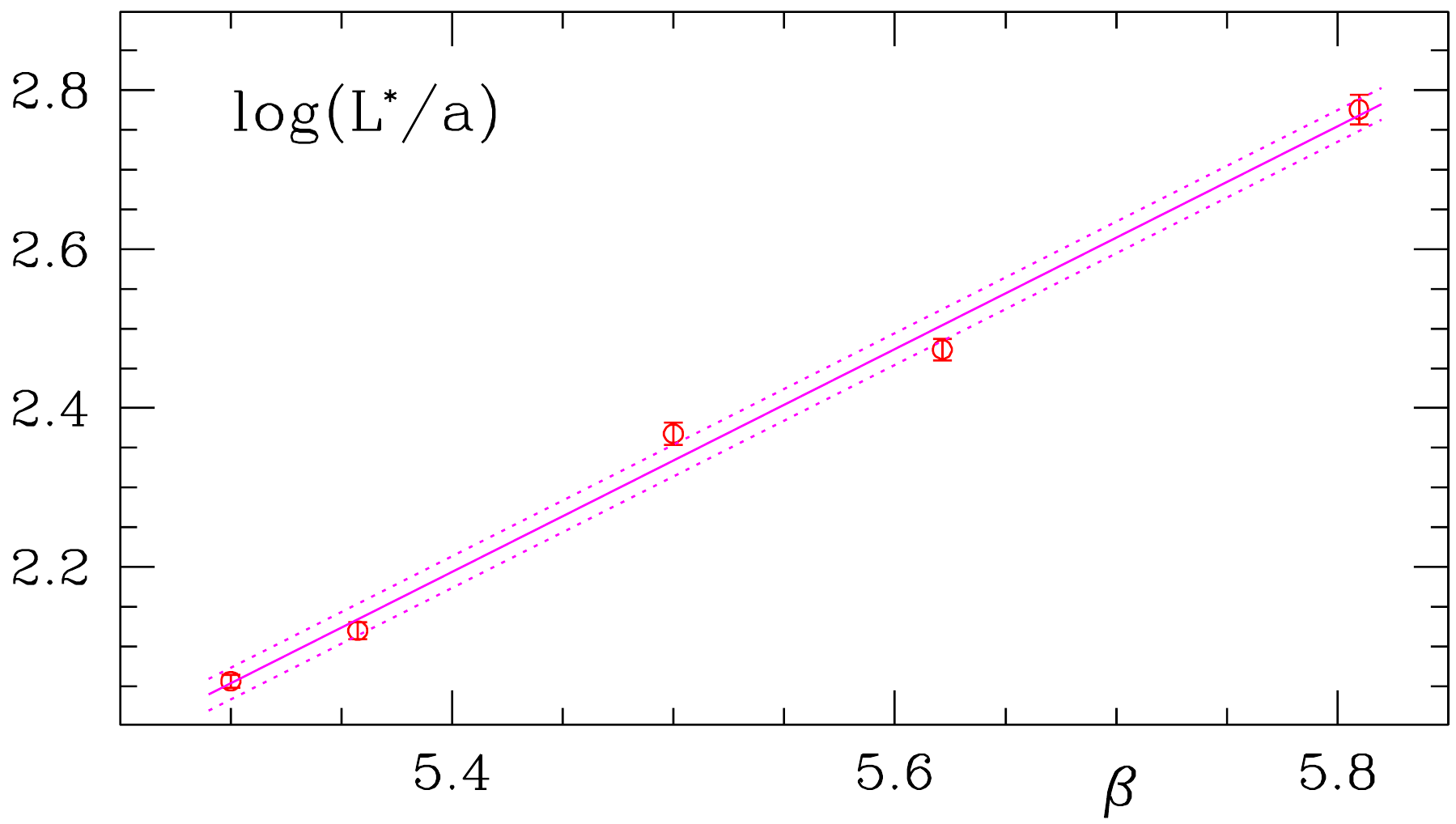}
\caption{\footnotesize
The results for $L^*/a$ as a function of $\beta$.
}\label{f:fit}
\end{center}
\end{figure}

Good guesses for the bare parameters $g_0,\kappa$ at a prescribed
$L/a$ are easily made
starting from table~11 of \cite{alpha:nf2}. When the result of 
a determination of $\gbar^2(L)$ is close to the target
\eq{e:lstar} and $m$ is close to zero, we may correct by a first order Taylor
expansion with derivatives \eq{e:ldep} and an estimate of 
\bes
  s = {1\over L} {\partial \over \partial m} \gbar^2|_{L}  \,.
  \label{e:zdep}
\ees
From the results at two different values of $m$ and
fixed $\beta=6/g_0^2=5.3574$ in \tab{tab:raw} we extract
\begin{wraptable}[13]{l}{0pt}
\small  \begin{tabular}{ll}
   \hline\\[-1.0ex]
   $\beta$ & $\log(L^*/a)$
\\[1.0ex]
   \hline\\[-1.0ex]
    5.3000   & 2.056(08) \\
    5.3574   & 2.120(11) \\
    5.5000   & 2.368(14) \\
    5.6215   & 2.474(14) \\
    5.8097   & 2.776(19) \\[1ex]
   \hline
  \end{tabular}
 \caption{Results for $L^*/a$.}
 \label{tab:res}
\end{wraptable}

\bes
  s= 2.2(5)\; \;\text{at }
  \gbar^2(L)\approx 5.5\,.
\ees
The rest of the simulation results of that table
are then corrected to match the target
with this value of $s$ (including its error) and with \eq{e:ldep}.
We arrive at \tab{tab:res} where a precision between 
0.8\% and 1.9\% is seen. These numerical values are very well
described by the simple linear interpolation formula
\be
  \log(L^*/a) = 2.3338 + 1.4025\,(\beta-5.5)
\ee
as seen in \fig{f:fit}
where a $\pm 0.02$ ``error band'' is shown.

%\onecol{3.7cm}{\input tabb.tex}
%\onecol{11cm}{
%%%%%%%%%%%%%%%%%%%%%%%%%%%%%%%%%%%%%%%%%%%%%%%%%%%%%%%%%%%%
%%%%%%%%%%%%%%%%%%%%%%%%%%%%%%%%%%%%%%%%%%%%%%%%%%%%%%%%%%%%%
%}

%%%%%%%%%%%%%%%%%%%%%%%%%%%%%%%%%%%%%%%%%%%%%%%%%%%%%%%%%%%%%%%%%%
\section{Outlook}

Using the estimate $a\approx0.08\,\fm$ at $\beta=5.3$
\cite{cern:I},
we have estimated the pairs $(\beta,L/a)=(5.5,32)$ and $(5.7,48)$
in order to remain in the large volume region $L\geq 1.9\,\fm$.
We are currently carrying out first simulations at these parameters.
Quark masses on the $L/a=48$ lattice are initially designed
to be only slightly below the mass of the strange quark. 
The reason is that our first goal is to carry out a precise scaling 
test, which is best done at not too small quark mass.
Combining with the results of 
\cite{cern:I,cern:II} a significant range of $a$ close
to the continuum can be covered.

The simulations are currently being done with the DD-HMC
algorithm~\cite{algo:L2}. Release 1.0 of Martin L\"uscher's 
software \cite{soft:DDHMC} has been adapted for the BlueGene/L
and an efficiency around 30\% has been achieved. The simulations
do thus run at a sufficient speed to expect results from
the BlueGene/L in J\"ulich rather soon. 
These efforts are part of coordinated lattice simulations (CLS)
carried out together with other lattice groups
at CERN, Madrid, Mainz, Rome (Tor Vergata) and Valencia.
\\[1ex]

\noindent
{\bf Acknowledgements.}
We thank NIC  for allocating computer time on the APE
computers to this project and the APE group for its help. This
work is supported by the  Deutsche Forschungsgemeinschaft
in the SFB/TR~09 and under grant
HE~4517/2-1, by
the European community through
EU Contract No.~MRTN-CT-2006-035482, ``FLAVIAnet''.

\bibliographystyle{JHEP}   %if you use h-elsevier.bst
\bibliography{refs}           %or whatever your .bib file i
% 
% \begin{thebibliography}{99}
%\bibitem{...} 
%....

%\end{thebibliography}

\end{document}